%% file: techRep.tex
\documentclass[12pt]{article}
\usepackage{epsfig,amssymb,amsmath,subfigure}
\newcommand{\dt}{\partial_t}

\newcommand{\da}{\partial_a}

\newcommand{\Amax}{a_{\mbox{\tiny{max}}}}
\newcommand{\Amin}{a_{\mbox{\tiny{min}}}}

\newcommand{\Pmin}{p_{\mbox{\tiny{min}}}}

\begin{document}
\def\trnum{2005-03}
\def\trauths{Bruce P. Ayati}
\def\trtitle{Modeling the Role of the Cell Cycle in Regulating {\it Proteus mirabilis} Swarm-Colony Development}
\input{trcover}

\baselineskip=.33truein
\bigskip
\title{Modeling the Role of the Cell Cycle in Regulating {\it Proteus mirabilis} Swarm-Colony Development}
\author{Bruce P. Ayati}
\date{}

\maketitle

\medskip
\begin{abstract}
We present models and computational results which indicate that the
spatial and temporal regularity seen in {\em Proteus mirabilis}
swarm-colony development is largely an expression of a sharp age of
dedifferentiation in the cell cycle from motile swarmer cells to
immotile dividing cells (also called swimmer or vegetative cells.)
This contrasts strongly with reaction-diffusion models of {\em
  Proteus} behavior that ignore or average out the age structure of the cell
population and instead use only density-dependent mechanisms.  We
argue the necessity of retaining the explicit age structure, and
suggest experiments that may help determine the underlying mechanisms
empirically.  Consequently, we advocate {\em Proteus} as a model
organism for a multiscale understanding of how and to what extent the life
cycle of individual cells affects the macroscopic behavior of a
biological system.
\end{abstract}



\section{Introduction}
\label{intro}
We present structured multiscale models and computational results of {\em Proteus mirabilis}
swarm-colony development which indicate that the spatial and temporal
regularity of the colony development is largely due to the regularity
of the cell cycle, namely a sharp age of dedifferentiation from motile
swarmer cells with a multinuclear structure to single-nucleus dividing
cells (also referred to as swimmer cells or vegetative cells elsewhere
in the literature).  We refer to our models as ``structured
multiscale'' because they link the cellular scale to the colony scale,
and represent the age dynamics of individual cells using a
population structure variable for age and an explicit aging term in
the differential equation.

This explanatory mechanism precludes the use of
reaction-diffusion models that either average over the age variable, such
as the models by Medvedev, Kaper, and Kopell \cite{MnKnK}, or ignore the age
structure from the outset, such as the models by Czir\'{o}k,
Matsushita, and Vicsek \cite{CnMnV}.

The necessity of explicitly retaining the cell cycle through an age
variable, rather than using density-dependent
processes alone, makes {\em Proteus} an ideal organism for a
multiscale understanding of how the kinetics at the individual cell level determines
the macroscopic behavior of a biological system.     

The models in this paper are based upon those presented in
\cite{JMB-proteus}.  The previous models assumed a sharp age of
dedifferentiation between motile swarmer cells and immotile dividing
cells.  This was incorporated into the model using an upper
bound on the age domain and then assuming that the swarmer-cell population
density was zero above this maximum age.  

The models in
\cite{JMB-proteus} used the critical insight of Esipov and Shapiro in
\cite{EnS} that the age dynamics of individual swarmer cells underly
the observed colony regularity and should be incorporated into the
model explicitly, using a population structure variable in the partial
differential equation for the swarmer-cell population.  The models in \cite{JMB-proteus} differed
substantially from those in \cite{EnS} by using a simpler diffusivity
which depended only on the swarmer-cell density rather than an
elaborate and unnecessary memory field -- one that proved to be insufficient
in generating the desired colony behavior when
accurate numerical methods were applied to the model equations.  The
models in \cite{JMB-proteus} also incorporated the known mechanism of
a density-dependent lag in differentiation between dividing cells and
swarmer cells.  Other differences included the use of a radially-symmetric geometry
rather than a linear geometry, diffusion based on isotropic random
motion rather than Fickian (symmetric) diffusion, and accurate
numerical methods.  The models in \cite{EnS}, when solved accurately, can
generate a regular temporal cycle and contain a control of the
ratio of swarm time to consolidation time, but are unable to generate
the regularity of the spatial structure, namely the bulls-eye pattern
with equally spaced concentric terraces.  An appendix in
\cite{JMB-proteus} contains a detailed discussion and computations of
the Esipov-Shapiro model, and the manuscript as a whole address in
more detail than here the differences between the two
modeling approaches.  The focus in this paper is the need to represent
the age structure of the swarmer-cell population explicitly, rather
than being able to use a reaction-diffusion framework to obtain an
understanding of the mechanisms that underly the observed macroscopic
behavior of {\em Proteus}.

This paper is organized as follows.  We provide a brief discussion of
{\em Proteus} biology, followed by a section going into more detail on
the critical differences and implications between the structured multiscale approach
used in this paper and the reaction-diffusion models used elsewhere.
We state the model equations and their biological meaning.  We then
discuss the numerical methodology used to solve the equations, and
present the results along with a discussion.  We close with our
conclusions and suggestions how the validity of the underlying assumptions of the
structured multiscale models and the reaction-diffusion models can be
determined empirically.


\section{{\em Proteus mirabilis} Swarm-Colony Development}
\label{biology}

When inoculated onto an agar surface, {\em Proteus} begin formation of strikingly regular spatio-temporal
  patterns that begin with three initial phases: a lag phase, a first
  swarming phase, and a first consolidation phase; these form what
  appear to be the center circle and first ring of a bulls-eye.  This
  is followed by repeating cycles of ring, or terrace, formation that
  consist of a swarming phase followed by a consolidation
  phase.  These subsequent terraces are equal in width and time of
  formation.  Perhaps most interestingly, the time of terrace
  formation is invariant under changes in the glucose or agar
  concentration of the substrate, although changes in agar
  concentration do affect the ratio of time spent swarming versus
  consolidating, and changes in glucose increase terrace width.
  Terrace formation does vary by temperature \cite{rauprich}. 

Varying glucose concentration greatly alters biomass production
without strongly affecting the spatial or temporal aspects of the swarm
and the consolidation cycle of terrace formation.  Changes in agar concentration do not change the
total time of terrace formation, but higher agar concentrations
shorten the time spent swarming and the width of the terraces, and
lengthen the time spent in consolidation \cite{EnS}.
 
A short summary of the physiology of {\em Proteus} cells can be found in a
section similar to this one in \cite{JMB-proteus}, and in more detail in
\cite{rauprich,WnSproteus}.
The critical aspect for understanding the macroscopic behavior is the
transition between the two cell types: mononuclear cells that are
motile in a liquid environment and immotile on an agar surface, called ``dividing'' cells in this paper
and ``swimmer'' or ``vegetative'' cells elsewhere in the literature;
and multinuclear filament cells which possess longer flagella and are
able to move on an agar surface by grouping together with other
swarmer cells in what are called ``rafts''.  The process in which dividing cells become
swarmer cells is called ``differentiation.''  Differentiation only
occurs above one dividing-cells density and below another.

When the multinuclear swarmer cells approach a maximum size, they rapidly
break down into mononuclear dividing cells.  This process is called
``dedifferentiation.''  Since mitosis results in exponential growth, the
size of a swarmer cell is an exponential weight times its age.
Thus, dedifferentiation occurs when swarmers reach a
maximum age.  

We suggest that the differentiation-dedifferentiation cycle of Proteus
cells, including a density-lag before differentiation occurs and a
sharp age near which dedifferentiation occurs, is a critical mechanism
for generating the macroscopic colony behavior, and thus
density-dependent mechanisms alone are insufficient.  

A deeper theoretical and empirical
understanding of how deviations from this cell cycle affect the
macroscopic behavior is needed to resolve the issue.  The models
presented in this paper aim to advance the theoretical understanding,
and experiments, perhaps similar to those by Hay, Tipper, Gygi, and Hughes
\cite{LrpProteus} where a mutated strain of {\em Proteus} was unable
to swarm due to a lack of sufficiently long flagella, can provide a
more sound empirical basis for the models.


\section{Critical Differences with Reaction-Diffusion Models}
\label{reaction-diffusion}

In this section we discuss the critical difference between the
structured multiscale models in this paper and reaction-diffusion
models of {\em Proteus} swarming.  We focus on the work by Czir\'{o}k,
Matsushita, and Vicsek \cite{CnMnV}.  This work makes explicit claims
of the sufficiency of a reaction-diffusion framework, utilizing only
density-dependent mechanisms, to answer all questions of interest
concerning {\em Proteus} swarm-colony development.  We direct less
attention to the paper by Medvedev, Kaper, and Kopell \cite{MnKnK}
because their work focused more on the interesting mathematical phenomenon
of a reaction-diffusion equation with periodic front dynamics and did
not make such claims about fully understanding the biology of
interest.  The Medvedev, Kaper, and Kopell \cite{MnKnK} paper was
discussed in more detail in \cite{JMB-proteus}.

The models in \cite{CnMnV} consist of a reaction-diffusion equation
for the swarmer cells and an ordinary differential equation at each
point in space for the ``vegetative'' cells.  The term ``vegetative''
cells was initially used in \cite{CnMnV} with the same meaning as ``dividing'' cells in
this paper, and later used to mean dividing cells and swarmer cells
which are unable to contribute to swarming.  The term ``consolidation''
has generally been used in the literature to mean the macroscopic-scale colony phase
when the front is not moving and the newly formed terrace is being
built up. Czir\'{o}k, Matsushita, and Vicsek use the term to mean the
cellular-scale process of
``dedifferentiation''.  This use is confusing given the differences in
scale, and that dedifferentiation
involves the breakup of multinuclear filament cells into several
mononuclear dividing cells.  

The models in \cite{CnMnV} use a minimum
and maximum population density of dividing cells for which
differentiation into swarmer cells can occur.  The upper limit has
been used in all models of {\em Proteus} swarm-colony development we
are aware of to date \cite{JMB-proteus,CnMnV,EnS,MnKnK}.  Medvedev, Kaper, and Kopell first saw the importance of the density lag in
differentiation from dividing to swarmer cells in the biological
literature \cite{WnSproteus} and incorporated it into a model.  The
lower limit is used here and in \cite{JMB-proteus,CnMnV}, but not in
\cite{EnS}.  One difference between the results presented here and in
\cite{JMB-proteus} and those in \cite{CnMnV} is that Czir\'{o}k,
Matsushita, and Vicsek get a loss of regularity in their results if
the density window for swarmer cell production becomes too wide.  This
loss of regularity takes the form of the consolidation phase becoming
shorter after each cycle.  In the models here and in
\cite{JMB-proteus}, the dividing-cell population density is scaled by
the density window for swarmer cell production in the
nondimensionalization.  The nondimensionalization as described in
\cite{JMB-proteus} shows why this is natural.  Instead we vary the
center of the interval, denoted by $v_c$.  As $v_c$ increases, so does
total cycle time and consolidation time, while swarm time remains
invariant.  Although we can remove the consolidation phase by changing
$v_c$, the result of each simulation retains the temporal and spatial
regularity.

The diffusivity used in \cite{CnMnV} is simpler than those used in
\cite{EnS} or \cite{MnKnK}, but is more complicated than the diffusivity used here
and in \cite{JMB-proteus}.

The models in this paper extend
those developed in \cite{JMB-proteus} to examine the
necessity of a sharp age of dedifferentiation from swarmer to dividing
cells.  The critical difference between the models in this paper and
in \cite{JMB-proteus} and those in \cite{CnMnV} is the mechanism of
dedifferentiation.  We argue that the assumptions underlying the
models are mutually exclusive.  The models in this paper require that
dedifferentiation occur almost exclusively near a critical age.  The
models in \cite{CnMnV} use a constant rate of dedifferentiation.

The important differences and commonalities between the age
explicit frameworks used here and those of Esipov and Shapiro \cite{EnS}
were discussed in much detail in \cite{JMB-proteus}.  A crucial
difference relevant to this discussion is that we dispense with a
memory field in the diffusion, which constitutes a built in
hysteresis.  Unfortunately, Czir\'{o}k,
Matsushita, and Vicsek \cite{CnMnV} focus on the memory field, rather
than the critical insight of Esipov and Shapiro of including the cell cycle in the model, in
their discussion of \cite{EnS}.  The major deficiencies
of the models and computations in \cite{EnS} are covered in much detail
in \cite{JMB-proteus}.  However, in the context of \cite{CnMnV}, this
paper and \cite{JMB-proteus} constitute a defense of the need to
explicitly represent age and the cell cycle in a model of {\em Proteus} swarm-colony development.

The models in \cite{JMB-proteus} reproduce
three major aspects of {\em Proteus} swarm-colony development:
temporal regularity of overall cycle time,
  spatial regularity, and control of the ratio of swarm time to
  consolidation time.  By retaining the explicit
  age structure in the mathematics, the models reproduce the colony behavior by
  relying on two main mechanisms: a density-lag in differentiation
  from dividing cells to swarmer cells, and a sharp age of
  dedifferentiation from swarmer cells to dividing cells.  The
  reaction-diffusion models of both
  in Czir\'{o}k, Matsushita, and Vicsek \cite{CnMnV} and Medvedev,
  Kaper, and Kopell \cite{MnKnK} reproduce only
  the temporal regularity of the overall cycle time and the spatial
  regularity.  No attempt is made to address the ratio of swarm time
  to consolidation time.  Given that this is controlled in the models
  in \cite{JMB-proteus} by an age threshold, it is not clear how such
  a control can be inserted into one of the reaction-diffusion
  modeling frameworks.  However, we concede that a control within a
  reaction-diffusion framework is within the realm of the possible.

Instead we argue the necessity of including the cell cycle
  explicitly by showing that moving from a sharp age of dedifferentiation
  to a broader distribution of dedifferentiation ages causes a
  breakdown in regularity, which would exclude density-dependent
  mechanisms alone from explaining {\em Proteus} swarm-colony and
  indicate the need to include age explicitly.  The models would be verified
  if experiments with strains of {\em Proteus} which do
  not have a sharp age of dedifferentiation give rise to irregular
  colony formation.


\section{The Models} 
\label{model}

The structured multiscale models in this paper are based on a mathematical
framework with some history behind it.  At the colony or population scale, Skellam \cite{skellam}
considered the effects of diffusion in his classic work
of 1951.  At the individual scale, Sharpe and Lotka in 1911 and McKendrick in 1926 considered population models with linear age
structure \cite{Webb}. More recently, Gurtin and MacCamy \cite{GnM1}
considered models with nonlinear age structure.  Rotenberg
\cite{roten} and  Gurtin \cite{gurtin} posed models dependent on both
age and space.  Gurtin and MacCamy \cite{GnM2} differentiated between
two kinds of diffusion in these models: diffusion due to random
dispersal, and diffusion toward an area of less crowding.  Existence
and uniqueness results can be found for various forms of these models
in  Busenberg and Iannelli \cite{BnI}, di Blasio \cite{diblasio}, di
Blasio and Lamberti \cite{DnL}, Langlais \cite{langlais1}, and MacCamy
\cite{maccamy}. Further analysis has been done by several authors
\cite{huang,KnL,langlais2,marcati}. 

We present in this section dimensionless equations, variables, and
parameters.  We assume the colony is radially symmetric and scale
space so that radius goes from zero to one.  We scale time and age by the time it takes a cell to
subdivide (typically 1.5 hours.)  A complete discussion of the nondimensionalization is contained in \cite{JMB-proteus}.
The dimensionless variables $r$, $a$, and $t$ represent radius in two-dimensional space,
age, and time, respectively.
The function $u(r,a,t)$ represents the swarmer-cell population density at
radius $r$, age $a$, and time $t$.  The functions $p(r,t)$ and $v(r,t)$
represent the biomass density of swarmer cells and the dividing-cell population
density, respectively, at radius $r$ and time $t$.  

The nondimensional model consists of the age-structured nonlinear diffusion system 
\begin{subequations}
\begin{eqnarray}
\dt u + \da u &=& \frac{1}{r}  \partial_r \big( r \partial_r ( D(p) u)
\big) - \mu(a)u, \  0 \leq r < 1, \, a>0, \, t > 0,
\label{swarm} \\
\dt v &=&  (1 - \xi(v))v + \int_0^\infty \mu(a) \, u \, e^{a} \ da, 0 \leq r \leq 1,  t > 0, \label{swim} \\
u(r,0,t) &=& \xi(v)v(r,t),  0 \leq r \leq 1, \ t > 0
\label{birth}
\end{eqnarray}
with boundary and initial conditions
\begin{eqnarray}
\partial_r \big( D(p(1,t)) u(1,a,t) \big) = 0,&&  a > 0 \ ,  t > 0, \\
u(r,a,0) = 0, &&  0 \leq r \leq 1, a \geq 0,
\label{initial-u}\\
v(r,0) = v_0(r),&&  0 \leq r \leq 1. \label{initial-v}
\end{eqnarray}
The total motile swarmer cell biomass is given by 
\begin{equation}
p(r,t) = \int_{\Amin}^{\infty} u(r,a,t) e^{a} \ da,  \qquad  0 \leq r \leq 1, \ t\geq 0. \label{Ps}
\end{equation}

{\em Proteus} move through a process of raft building that requires
two things: sufficient maturity in swarmer cells to contribute
to raft building, and a sufficient biomass of mature cells to form the
rafts. The lower limit of integration, $\Amin$, is the minimum age when a
swarmer cell is sufficiently large, with sufficiently long flagella,
to contribute to motion on the agar surface.  This
parameter controls the ratio of swarm time to consolidation
time without changing the total time of ring formation.  The parameter
$\Amin$ is related to agar concentration.  The higher the
concentration, the drier the surface, raising the value of $\Amin$.
Higher agar concentration, and thus higher $\Amin$, shortens
the swarming phase and lengthens the consolidation phase in terrace formation.

The diffusivity has the form 
\begin{equation}
  D(r,t) = D_0 \max\big\{ (p(r,t)-\Pmin), 0\big\}. \label{diffusivity}
\end{equation}
The parameter $\Pmin$ is the minimum biomass needed for swarmers to
build rafts capable of moving on the agar surface.

We use a differentiation function with a lag phase that is a $C^1$ piecewise cubic with
support of length 2: 
\begin{equation}
 \xi(v) =  \left\{ \begin{array}{ll} \xi_0 \left( 2
                      |v-v_c|^3 - 3(v-v_c)^2 + 1 \right), &
                      |v-v_c| \leq 1, \\
                      0, & \mbox{otherwise,}
                      \end{array} \right.  \\ \label{diff}
\end{equation} 
The interval $[v_c-1,v_c+1]$ is the
swarmer-cell production window.  The interval width of 2 was obtained
in the nondimensionalization.  The parameter $v_c$ represents the
lag phase in swarmer-cell production \cite{MnKnK}.  We use a compactly
supported cubic function because it is smooth.  The important aspects
of the functional form of $\xi$ are that it is zero above and below
certain thresholds.  An appendix in \cite{JMB-proteus} shows
that the nature of the results do not depend strongly on the shape of
the curve.  The shape of the curve does help with numerical issues
concerning the degenerate diffusion.  The parameter $v_c$ has no
analog in \cite{EnS}.

We take the initial condition to be 
\begin{equation}
 v_0(r) = \left\{ 
                \begin{array}{ll}
                      v_h \left( 2\left( \frac{r}{r_0} \right)^3
                            - 3\left( \frac{r}{r_0} \right)^2 + 1 \right), & 0 \leq r \leq r_0, \\
                      0, & r > r_0.
                \end{array} 
          \right.
\label{initial-form}
\end{equation} 

Again, it is mass, not shape, that matters qualitatively. Shape effects numerical efficiency.

The major difference in the models in this paper and
those in \cite{JMB-proteus} is the dedifferentiation modulus, $\mu(a)$, and
the resulting behavior of the system to changes in dedifferentiation from
swarmer to dividing cells.  In \cite{JMB-proteus}, we used a form that
was the sum of a Heaviside function and a Gaussian distribution so
that in the limit of the spread parameter $\sigma \rightarrow 0$ it
would be transparent that we obtain a model with no explicit function
$\mu$ but could instead represent the sharp age of dedifferentiation
by an upper limit in the age domain.  Only the limiting case was
treated in \cite{JMB-proteus}.

Our goal in this paper is to show the necessity of retaining explicit
space structure and a sharp age of dedifferentiation.  Consequently,
we examine the change in swarm-colony regularity as dedifferentiation
become less sharp.   We use a function with similar shape but a
simpler expression than the dedifferentiation modulus used in
\cite{JMB-proteus},
\begin{equation}
 \mu(a) = \frac{\mu_o}{\sigma} \left( \tanh\left(
     \frac{a-\Amax}{\sigma}\right) + 1 \right), \label{mu}
\end{equation}  
\end{subequations}
where $\sigma$ is the spread parameter and $\mu_0$ is a height
parameter.  This function is equivalent to a Fermi distribution.  For
a fixed $\mu_0$ and taking the limit $\sigma 
\rightarrow 0$, we get a situation where swarmer cells 
dedifferentiate into their component dividing cells all at the same
age, $\Amax$.  For $\sigma$ small but not equal to zero, this dedifferentiation modulus represents
 a situation where the probability of dedifferentiation is low for
 young swarmers, increases rapidly as they approach $\Amax$ in age, and
 remains high afterwords.  What is important is the rapid change in
 the dedifferentiation rate.  Simulations carried out with a piecewise
 continuous linear function $\mu$, that is zero below
 $\Amax-\sigma$, $\mu_0$ above $\Amax+\sigma$, and increases linearly
 in between, show no qualitative and minor quantitative differences
 with the form in equation (\ref{mu}). 

\begin{figure}[t]
\centering
\subfigure[Sharp age of dedifferentiation]{
\epsfig{file=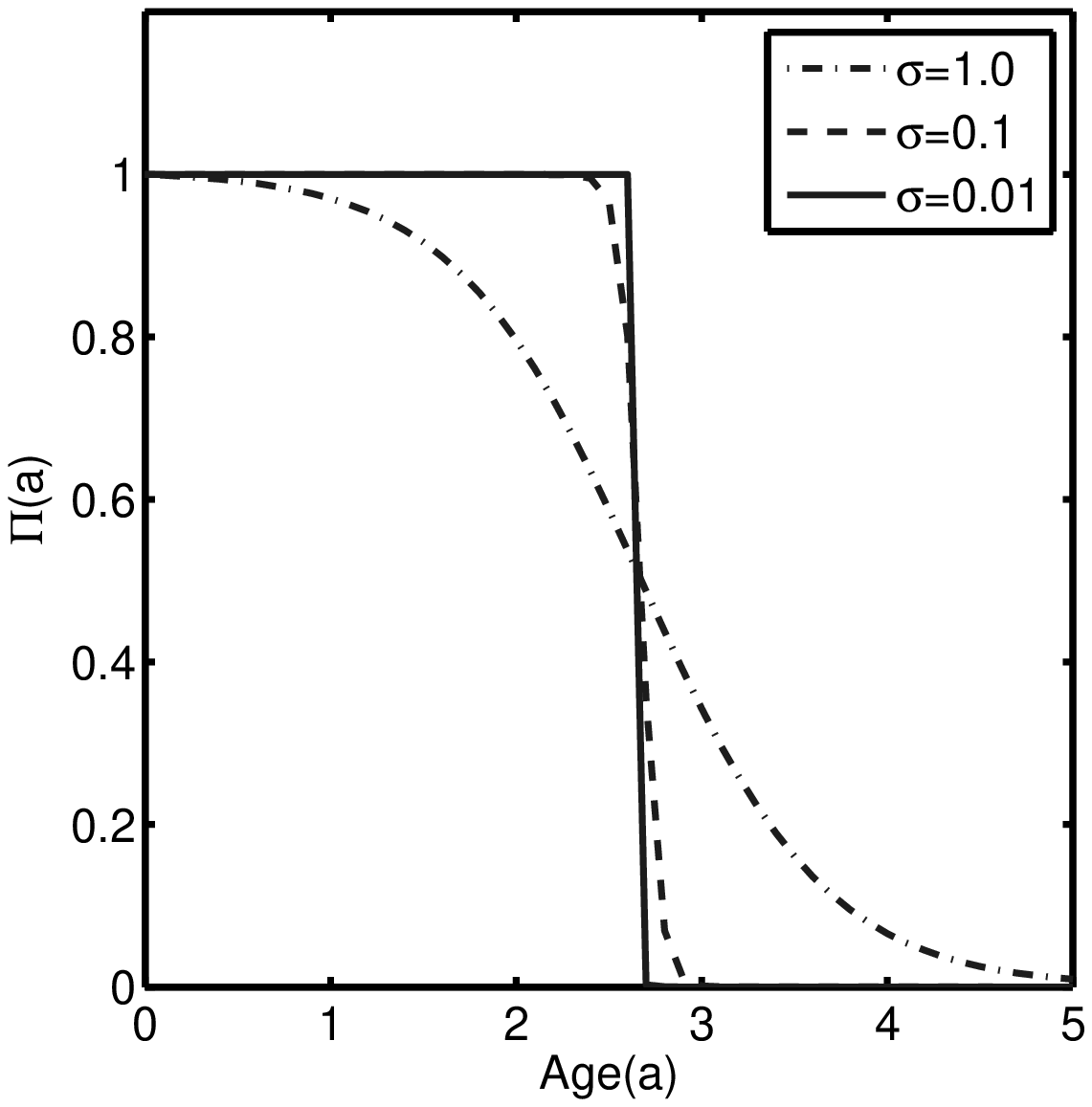,height=3in}
\label{PiHeavi}
}
\subfigure[Non-sharp age of dedifferentiation]{
\epsfig{file=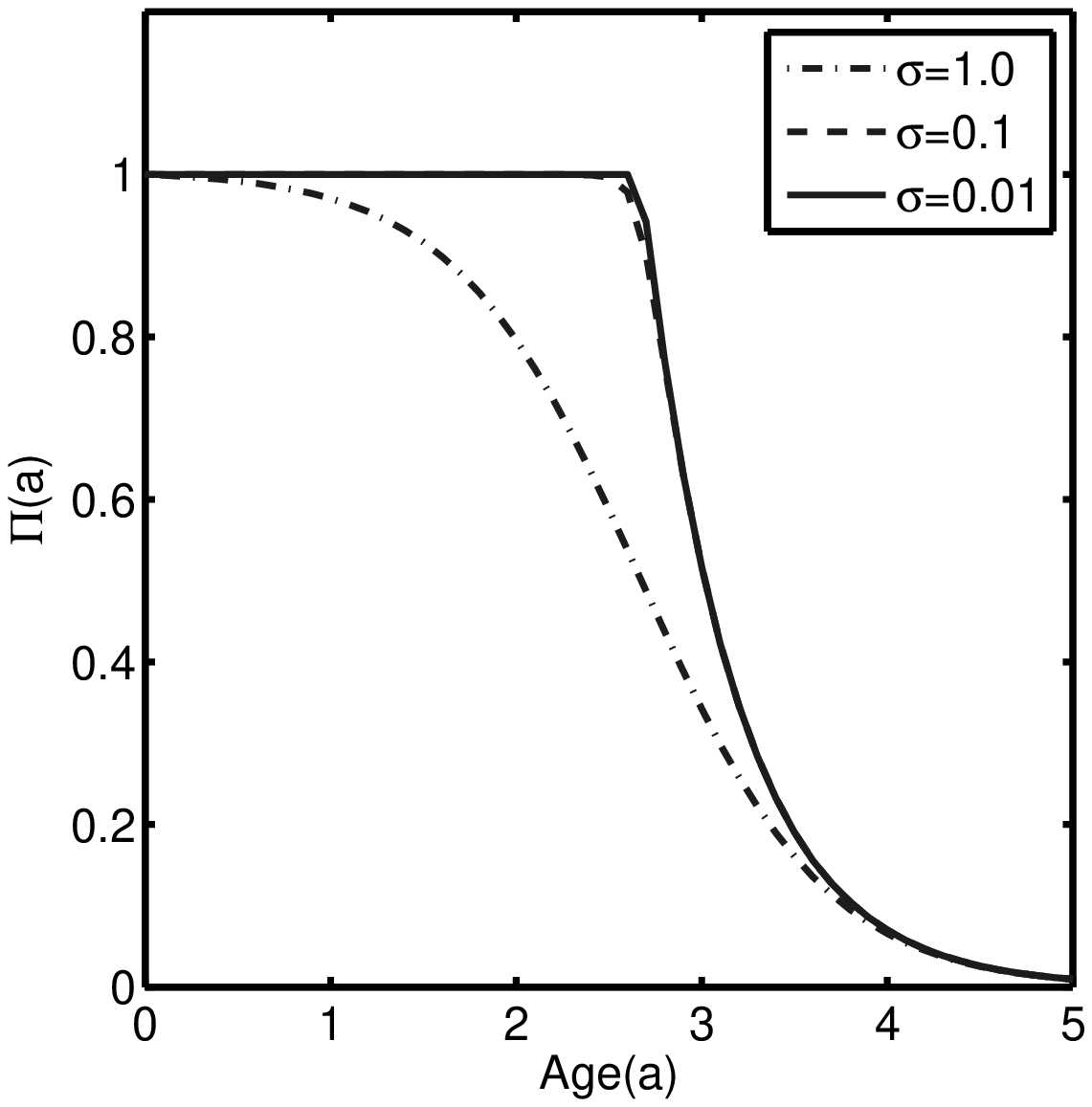,height=3in}
\label{PiDecay}
}
\caption{The probability $\Pi(a)$ that a swarmer cell has not dedifferentiated by age
$a$.  In figure \ref{PiHeavi}, $\mu_0=1.0$ in equation
(\ref{PiSpecific}).  In figure \ref{PiDecay}, $\mu_0=\sigma$ in equation
(\ref{PiSpecific}). These figures illustrate the need to
increase the magnitude of the dedifferentiation modulus to
approximate a sharp age of dedifferentiation from swarmer cell to
dividing cell in the models.}
\label{PiFigure}
\end{figure}

It may be unclear at first why the magnitude coefficient of $\mu$ (the ratio
$\mu_0/\sigma$) should increase as the hyperbolic tangent approaches a step function.
The intent is to model the effect on colony behavior of moving away or
toward a sharp age of dedifferentiation in the model.  We show that if we keep
$\mu_0/\sigma$ fixed as we vary $\sigma$, we converge to something
other than a sharp age of dedifferentiation as $\sigma \rightarrow 0$.
The probability that a swarmer cell has not dedifferentiated by age
$a$ is given by 
\begin{equation}
\Pi(a) = \exp \left(-\int_0^a \mu(\omega) \ d\omega \right)
\label{Pi}
\end{equation}
(see pg. 82 of \cite{cushing}.)  Using the specific form of $\mu$ in equation
(\ref{mu}), we obtain
\begin{equation}
\Pi(a) = \exp\left(\frac{-\mu_0 a}{\sigma}\right) \left(\frac{\cosh \frac{a-\Amax}{\sigma}}{\cosh \frac{-\Amax}{\sigma}} \right)^{-\mu_0}.
\label{PiSpecific}
\end{equation}
Figure \ref{PiFigure} illustrates how keeping $\mu_0$ fixed as $\sigma
\rightarrow 0$ yields convergence of $\Pi$ to a step function, whereas
keeping $\mu_0/\sigma$ fixed gives convergence to a function with decay to zero rather than a discontinuity.

The correct way to represent a sharp age of dedifferentiation has proven to be
nontrivial.  Models in \cite{EnS} (specifically ``Model A'' as defined
on pp. 252-253) conflate the desired probability
density function (pdf) for dedifferentiation ages (a Dirac delta function
centered at the sharp age of dedifferentiation) with the
dedifferentiation modulus.  As can be seen by setting
$\mu(\omega) = \delta(\omega-\Amax)$ in equation \ref{Pi}, a Dirac
delta function as the dedifferentiation modulus would give rise to a
situation where there is no dedifferentiation until $\Amax$, at which point the population is reduced by a factor of
$1/e$.  This remaining swarmer-cell population never dedifferentiates, and continues to age (and hence grow),
resulting in no consolidation period whatsoever. 

Esipov and Shapiro \cite{EnS} attempted to examine the importance of a sharp age
of dedifferentiation by treating two models, ``Model A'' with
dedifferentiation as mentioned above, and a ``Model B'' with a
constant dedifferentiation modulus.  As discussed in
\cite{JMB-proteus}, ``Model A'', when solved accurately, does not
reproduce the observed {\em Proteus} colony behavior, which limits the
utility of that treatment.  Just as important, only considering the
extreme cases leaves open several questions.  All swarmer cells do not
dedifferentiate at {\em precisely} the same age; this is a construct
useful for modeling.  Does even a minor
deviation from a sharp age of dedifferentiation lead to colony
irregularity?  If so, this would invalidate the model.  Also left open
is the nature of the transition from regularity to irregularity of
colony formation.  We attempt to answer these question using
mechanisms in our dedifferentiation function absent from those in \cite{EnS}.


\section{Computation and Results}
\label{results}

In this section we briefly describe the computational methods used to
obtain our solutions and then present those solutions with their
biological interpretations.

The computational methods and software used in this paper are
essentially those used to solve the model equations in
\cite{JMB-proteus}.  The main exception is
the use of a modulus, $\mu$, rather than a finite
age domain truncated at the septation age, to model dedifferentiation
from swarmer to dividing cells.  This
required changes in how information was passed from the dividing-cell
equation to the swarmer-cell equation.  Previously, the population of
cells undergoing dedifferentiation could be determined from the oldest
age cohort at each point in space at the end of the time step and
passed to the dividing-cell equation at the start of the next time
step.  For the models in this paper, the population of
dedifferentiating swarmer cells had to be computed across all age cohorts at
each point in space and then
aggregated for passing to the dividing-cell equation.  

A more detailed description of the numerical methodology as specifically
applied to {\em Proteus} can be found in \cite{JMB-proteus}, and the
original formulations and analyses can be found in
\cite{age-pwconst-paper,age-general-paper,step-doubling-paper}.  To
summarize, the
software uses a moving-grid Galerkin method in the age variable to
decouple age from time, which results in a degenerate
nonlinear parabolic partial
differential equations in space and time for each age cohort.  The
parabolic system is then solved using a step-doubling extrapolation
method \cite{step-doubling-paper}.

We emphasize the importance of reliable numerics.   In the systems
presented in this paper and in \cite{JMB-proteus}, the regularity in terrace
formation is due to the regularity in the aging of the swarmer cells.
A coarse, stage-structured numerical approximation to the continuous
aging term, such as that used in \cite{EnS}, can constitute a
qualitatively different swarmer-cell life cycle than that of the
original continuous model. Also, the degenerate diffusion can be
altered by the choice of the time step.  The numerical computation may then show
periodic front dynamics that are not obtained by an accurate solution
of the original continuous equations, but are induced by a regularity
in the numerics.  An appendix in \cite{JMB-proteus} details
how accurate computations change the behavior of the
Esipov-Shapiro system from what is presented in \cite{EnS}.

A uniform spatial discretization of size $\Delta x = 1/300$, and a
uniform age discretization of $\Delta a = 1/40$, using piecewise
constant basis functions, was sufficient to
solve the system within a relative error in the $L^2$-norm of less than
1\%.  A uniform age discretization in the context of a moving grid
method means that all but the first and last age intervals are
constant in length, and that a new age interval is introduced at the
birth boundary when the old birth interval reaches $\Delta a$ in length.
Convergence in time was obtained by adjusting a tolerance parameter
for the adaptive time-stepping in the step-doubling algorithm so that
the relative error in the $L^2$-norm was also less than 1\%.

The goal of this paper is to model the effects of a non-sharp
dedifferentiation age on colony regularity.  Consequently, we focus on
the behavior of the system defined by equations (\ref{swarm})-(\ref{mu}) as we vary $\mu_0$ and $\sigma$.  For the computations in
this paper, we set $\Amax = 2.67$, $v_c = 8.0$, $D_0 = 2\times 10^{-3}$,
  $\Pmin = 0.5$, $\xi_0 = 0.5$, and $\Amin=0$. 

The
response of the system to changes in these six parameters was
investigated in detail in \cite{JMB-proteus} (we have dropped the
``hat'' or capital notation in this paper, which was used in \cite{JMB-proteus}
to denote the nondimensional version of a variable or parameter.)  In
summary, $\Amax$ is the nondimensional sharp age of dedifferentiation (in this paper
its meaning changes to become the ``center'' of a mollified step
function.)  As $\Amax$ increases, so does the time spent swarming, the
terrace width, and the overall cycle time.  The time spent in
consolidation decreases.  The parameter $v_c$ is the nondimensionalized
lag in swarmer cell production, $\xi$. As $v_c$ increases,
consolidation time, total cycle time, and terrace width increase,
where as swarming time remains nearly invariant.  The constant of
diffusion, $D_0$, represents changes in glucose concentration and has
a significant effect only on terrace width.  The parameter $\xi_0$ is
the constant of the differentiation ratio $\xi$, and lies between zero
and one.  As $\xi_0$ increases, so does swarming time and terrace
width, whereas total cycle time and consolidation time decrease.

The parameter $\Amin$ is
the minimum age at which a swarmer cell can contribute to raft
building and movement and controls the ratio of swarm time to
consolidation time within a cycle.  It only has meaning in the context
of an age-explicit model.  As agar concentration increases,
so does the difficulty of moving on the substrate, which is
represented by increasing $\Amin$.  As $\Amin$ increases, swarming
time and terrace width decrease, consolidation time increases, and
total cycle time remains the same.

The parameter $\Pmin$ is
the minimum swarmer biomass needed for raft formation and the onset of
swarming.  It is thus measured in the same units as population
density, and differs in an
important respect from $\Amin$, which is measure in the same units as
age.  This difference is worth noting given that we argue the
necessity of including both density-dependent processes and the cell
cycle in a model, rather than relying on density-dependent mechanisms alone to
explain the observed {\em Proteus} behavior.  The response of the
system to changes in $\Pmin$ differs from changes to $\Amin$.  As $\Pmin$ increases,
swarming time, total cycle time, and terrace width decrease,  whereas
consolidation time increases.  The parameter $\Pmin$ is needed to
create a distinct consolidation phase.

\begin{figure}
\subfigure[Simulated Colony]{
\epsfig{file=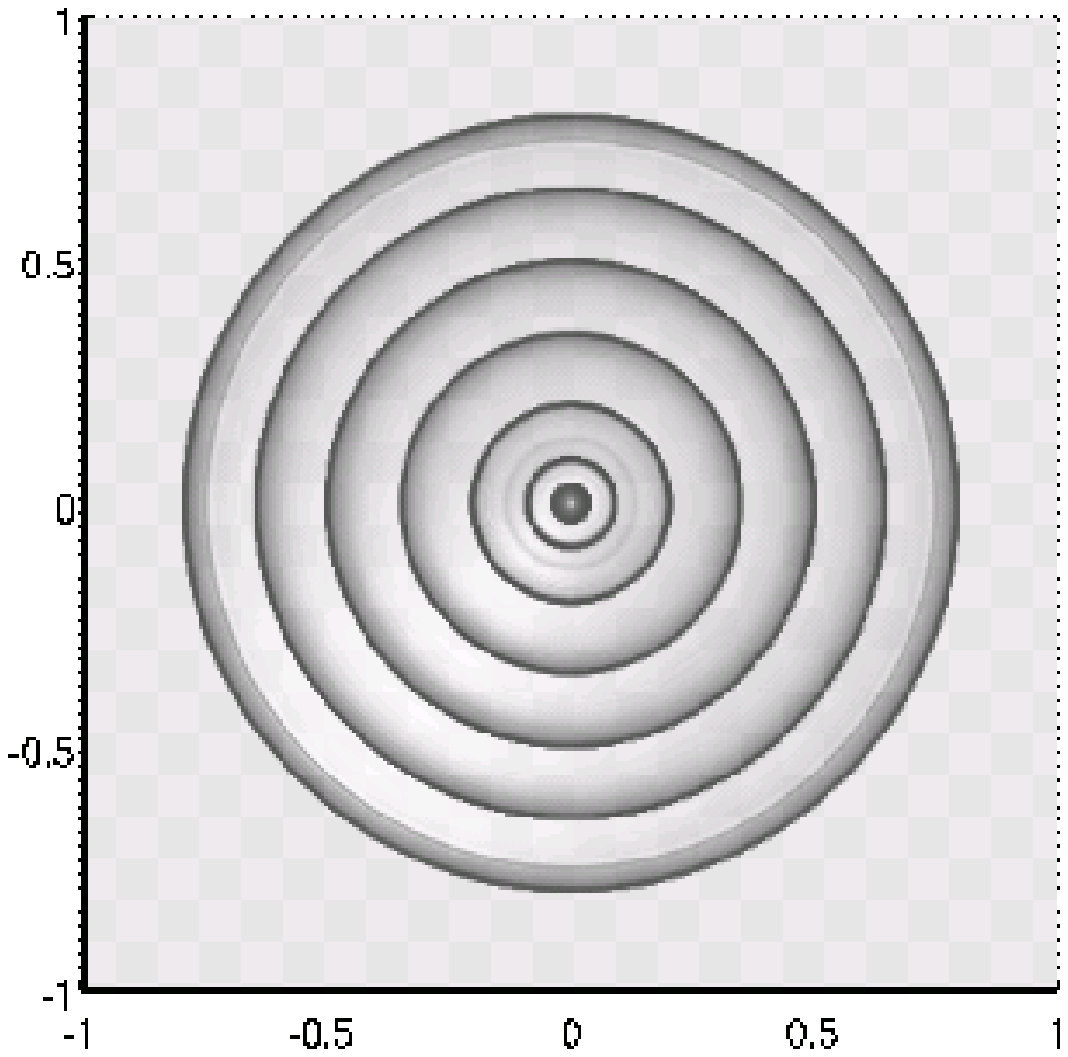,height=2in}
\epsfig{file=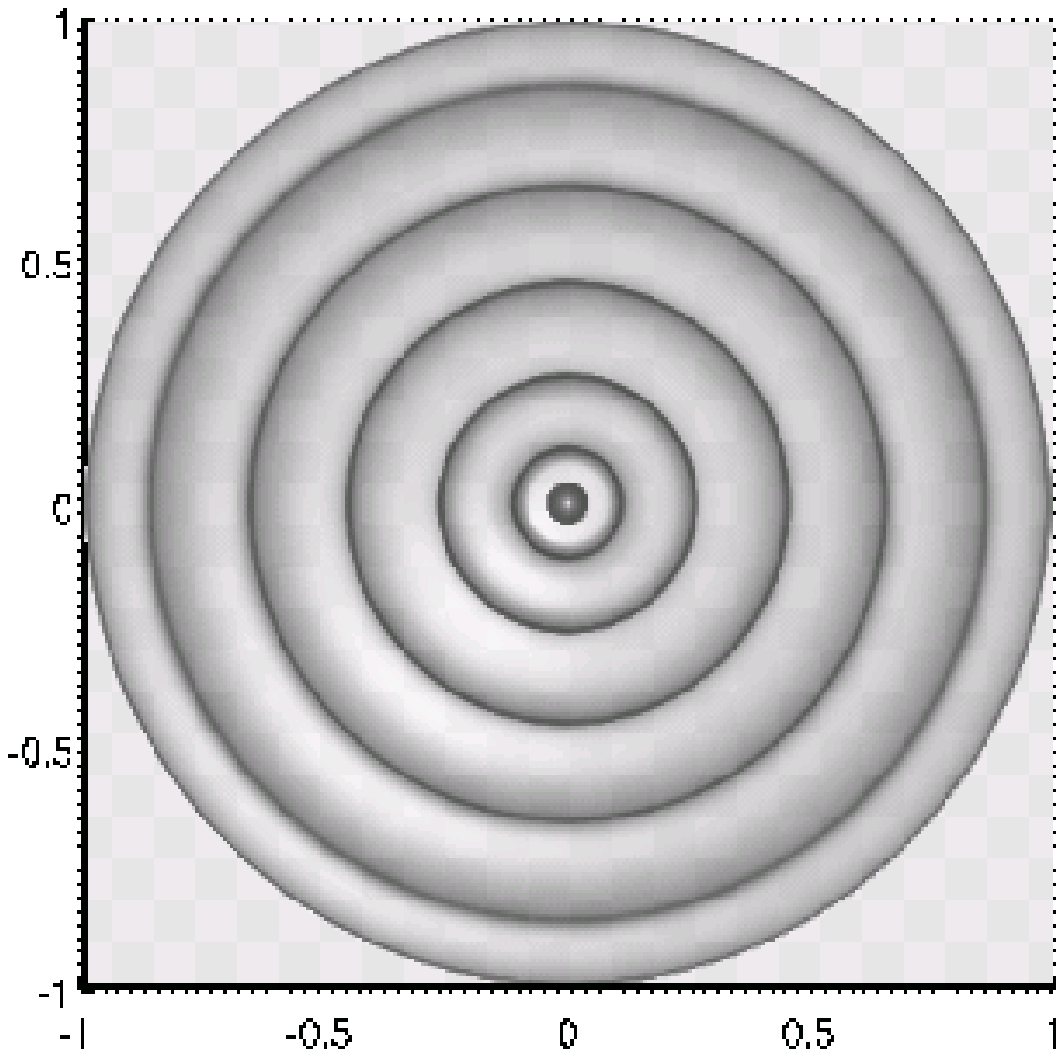,height=2in}
\epsfig{file=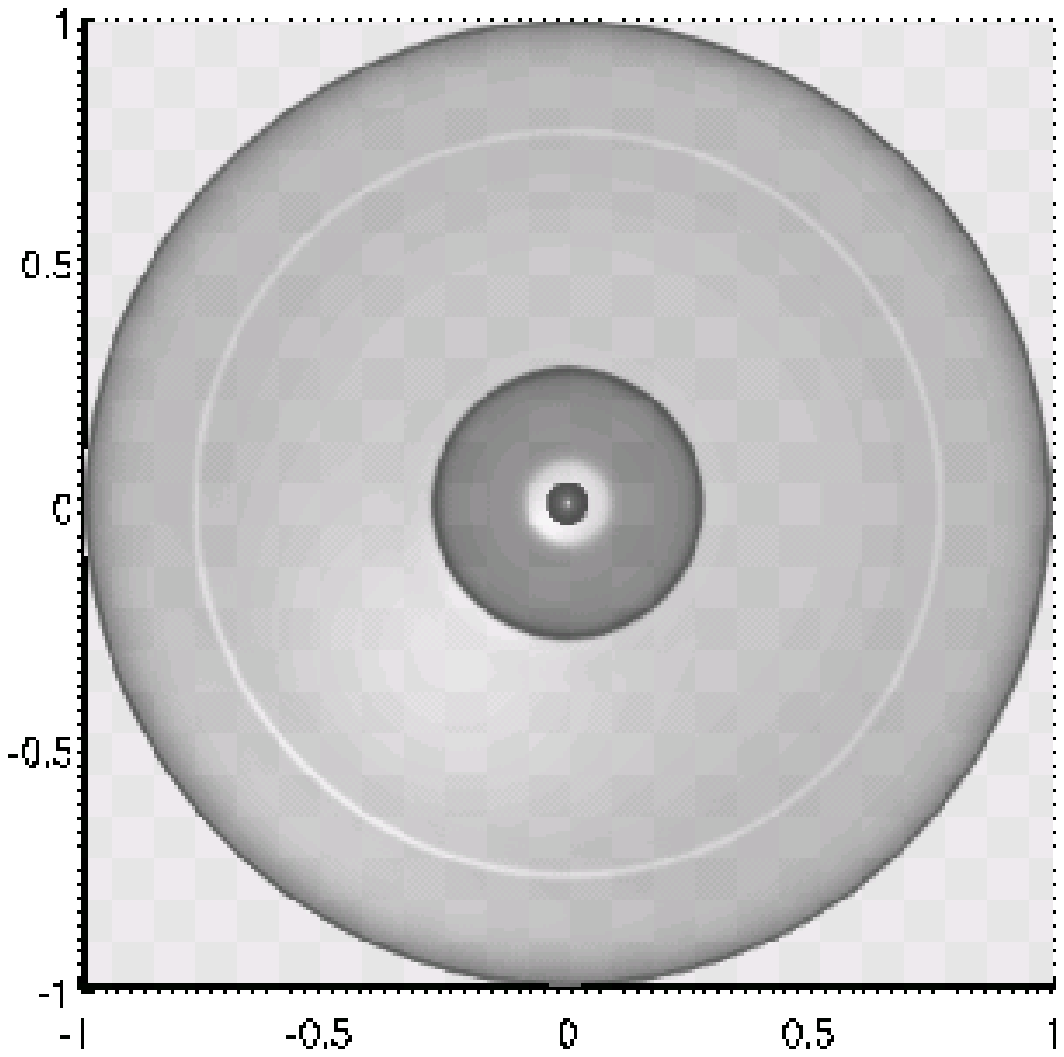,height=2in}
\label{colonies}
}
\subfigure[Position of Colony Front]{
\epsfig{file=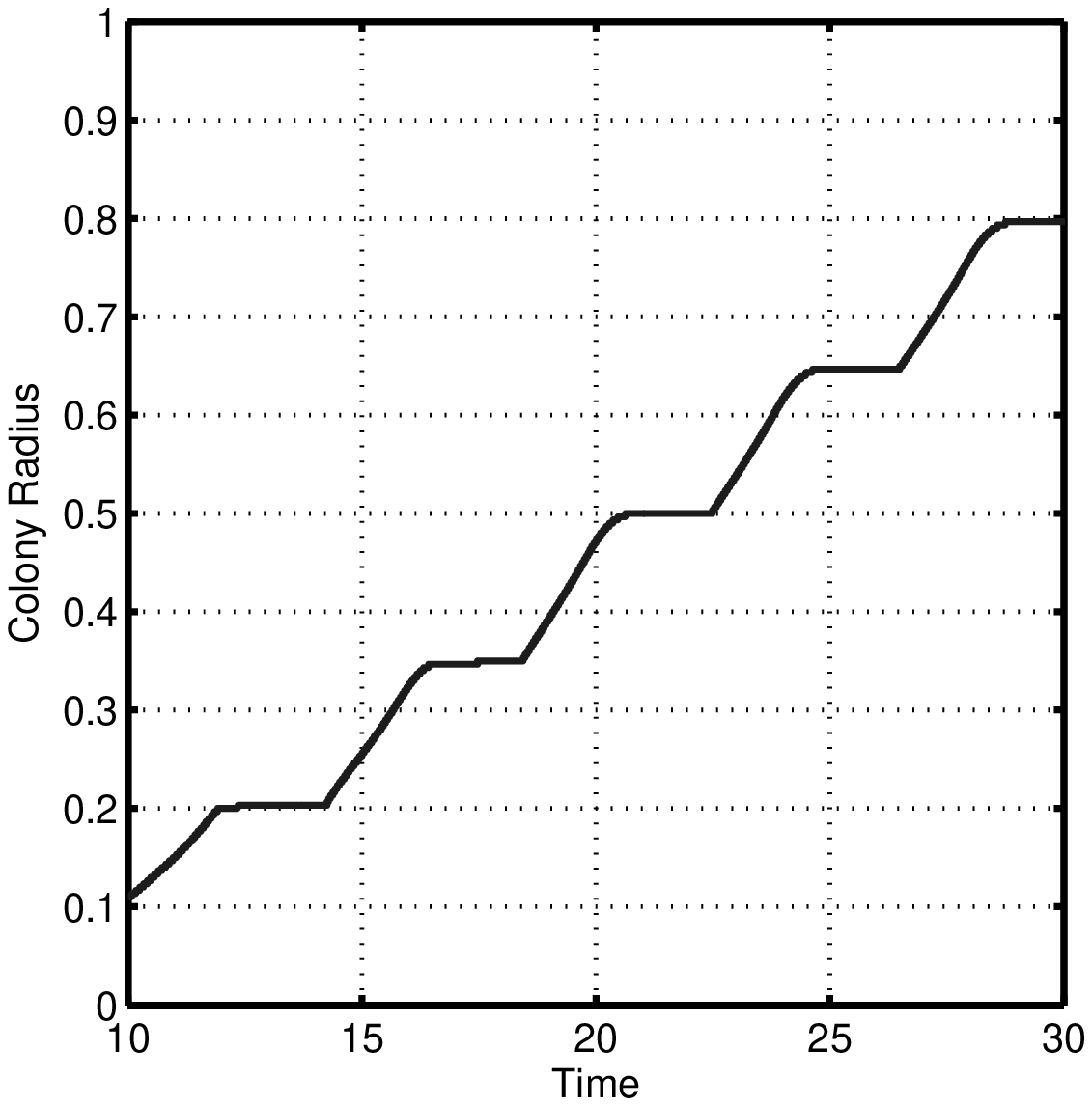,height=2in}
\epsfig{file=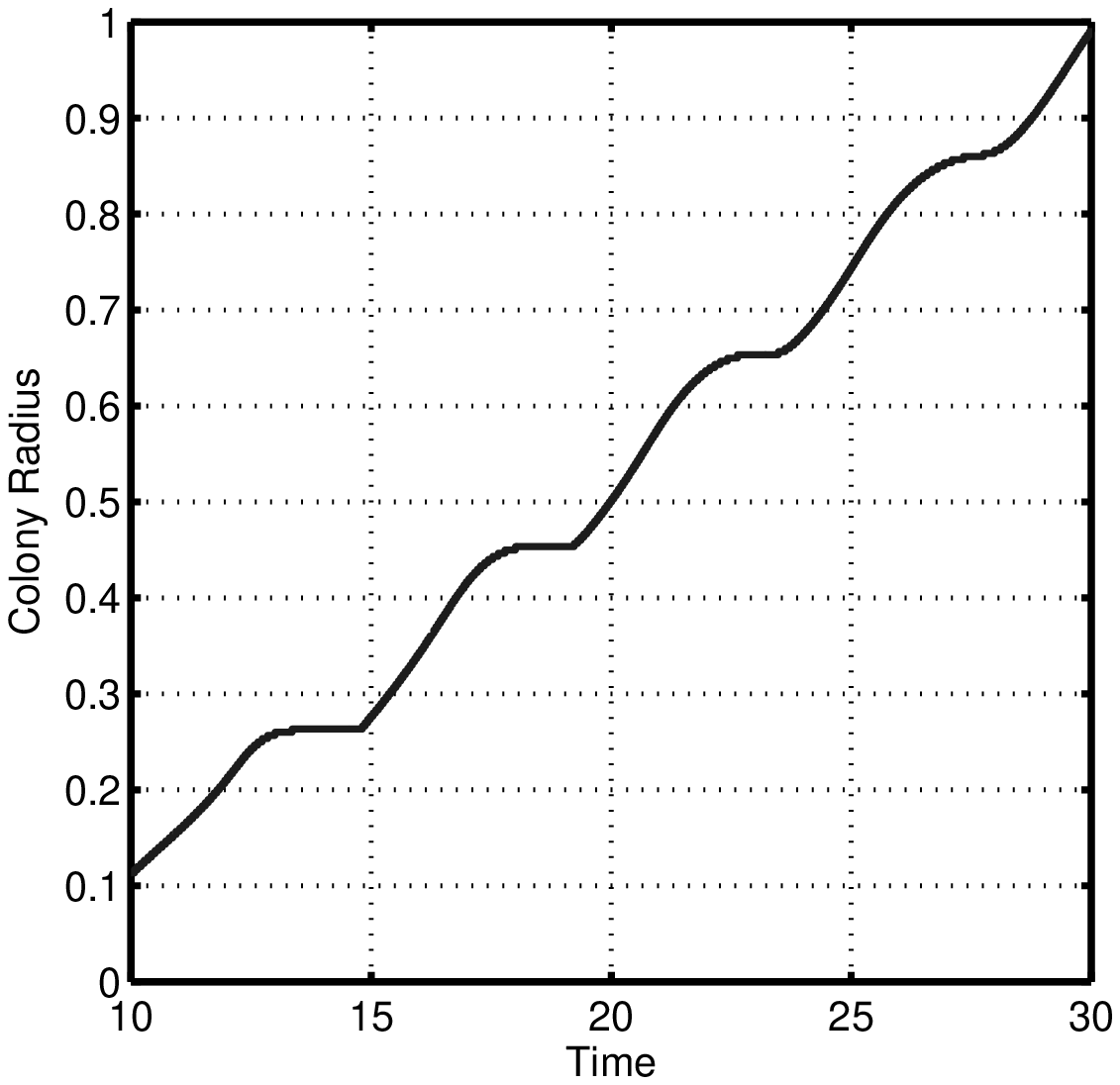,height=2in}
\epsfig{file=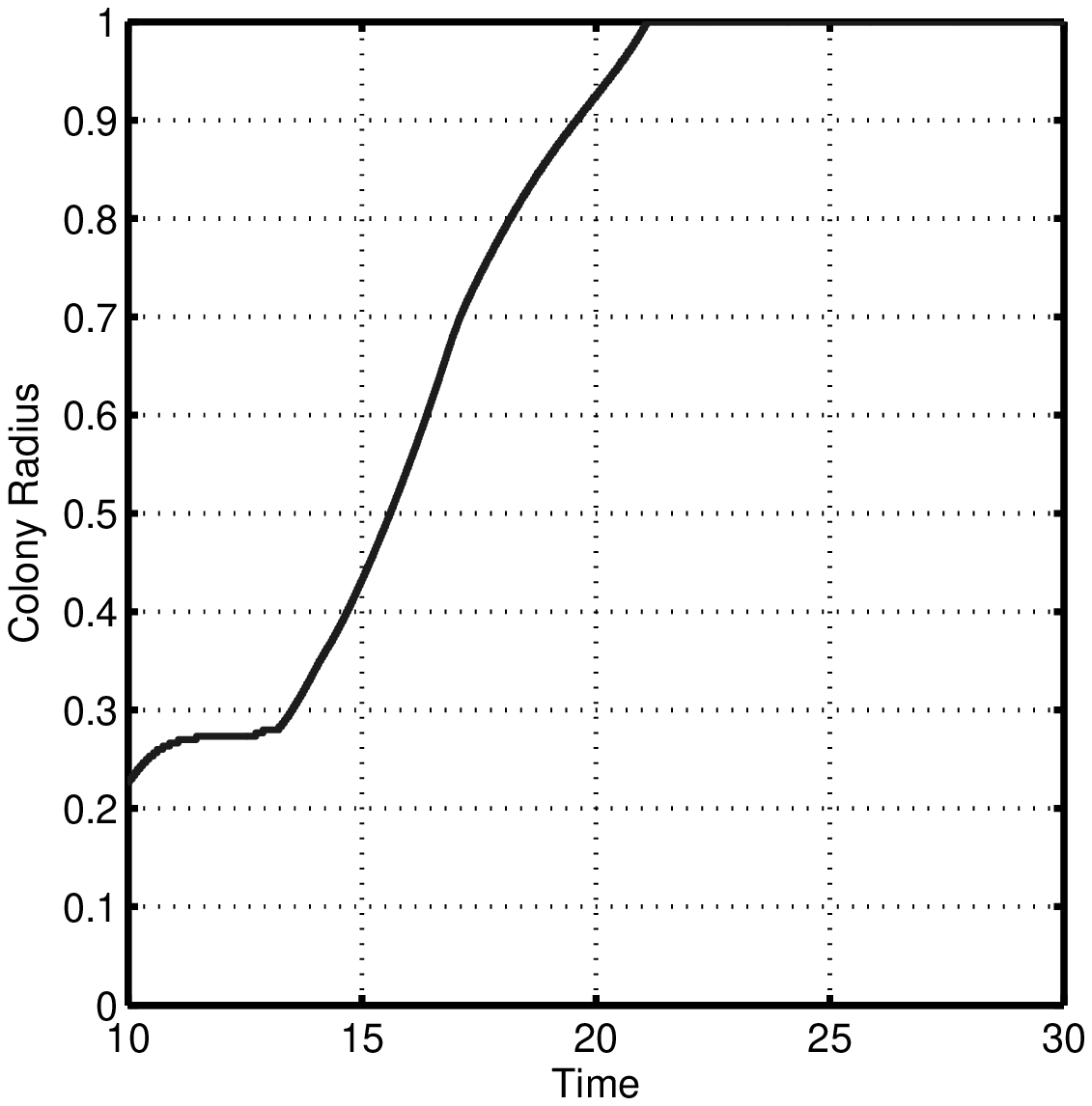,height=2in}
\label{fronts}
}

\caption{Computed {\em Proteus mirabilis} swarm colonies showing a
  breakdown of spatial and temporal regularity as the
  dedifferentiation function becomes less sharp. Note that temporal
  regularity breaks down before the apparent
  spatial regularity.  Figure \ref{colonies} consists of the 3D plots of
  the swarmer- and dividing-cell biomass viewed from directly above
  with a combination of ambient light, diffuse reflection, and specular
  reflection. The time of the snapshot is the earlier of $t=30$ or
  when it hits the boundary.  Figure \ref{fronts} shows the corresponding radius of the swarm
  colonies as functions of time.  Areas where the function has zero
  slope correspond to consolidation periods in the swarm-colony
  development.  The dedifferentiation parameter $\mu_0$ varies for the three
  colonies, left to right, from $\mu_0=0.1$ to $\mu_0=0.01$ to
  $\mu_0=0.001$. Parameters common to all three swarm colonies are
  $\sigma=0.01$, $\Amax = 2.67$, $v_c = 8.0$, $D_0 = 2\times 10^{-3}$,
  $\Pmin = 0.5$, $\xi_0 = 0.5$, and $\Amin=0$. } 
\label{colonyPlots} 
\end{figure}

To understand the breakdown of colony regularity as the age of
dedifferentiation become less sharp, we examine first the response to
changes in $\mu_0$.  Decreasing $\mu_0$ for a fixed $\sigma$ has the
effect of decreasing the amount of dedifferentiation after $\Amax$,
which can ultimately result in a continuation of swarming and a loss
of a consolidation phase.  As shown in figure \ref{colonyPlots}, this
breakdown can be quite irregular.  The temporal breakdown occurs
before an apparent breakdown in spatial regularity.  Not shown in the
figure is the effect of increasing $\mu_0$.   Increasing $\mu_0$ above
what is shown results first in regular colonies with narrow terrace
widths, and then the loss of regularity through the loss of any meaningful
swarming phase.  This is intuitive since increasing $\mu_0$ results in
earlier and more frequent dedifferentiation, which in turn results in
a lack of large swarmer cells to contribute to raft building and the
consequent swarming.

\begin{figure}[t]
\centering
\subfigure{
\epsfig{file=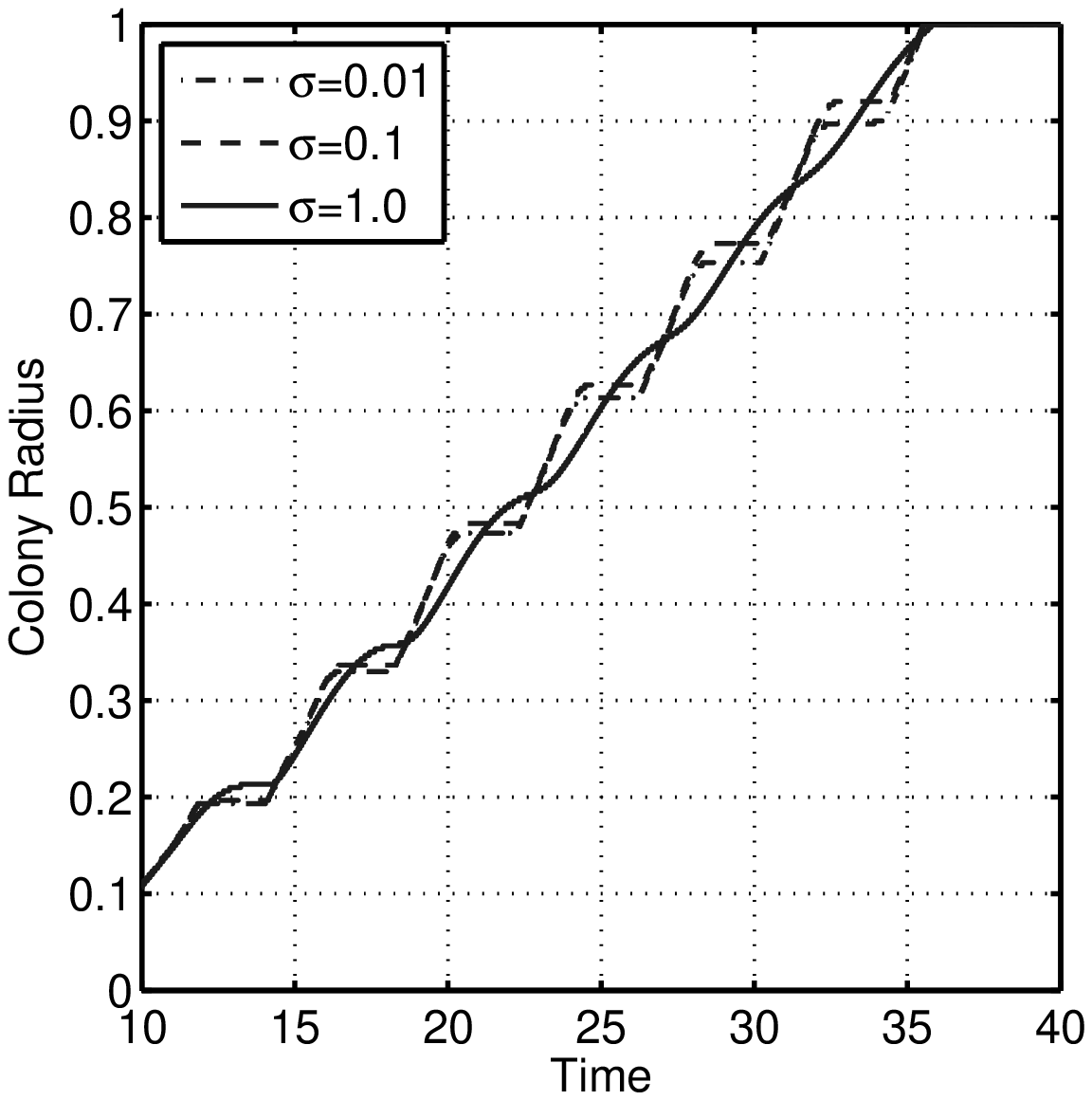,height=3in}
}
\subfigure{
\epsfig{file=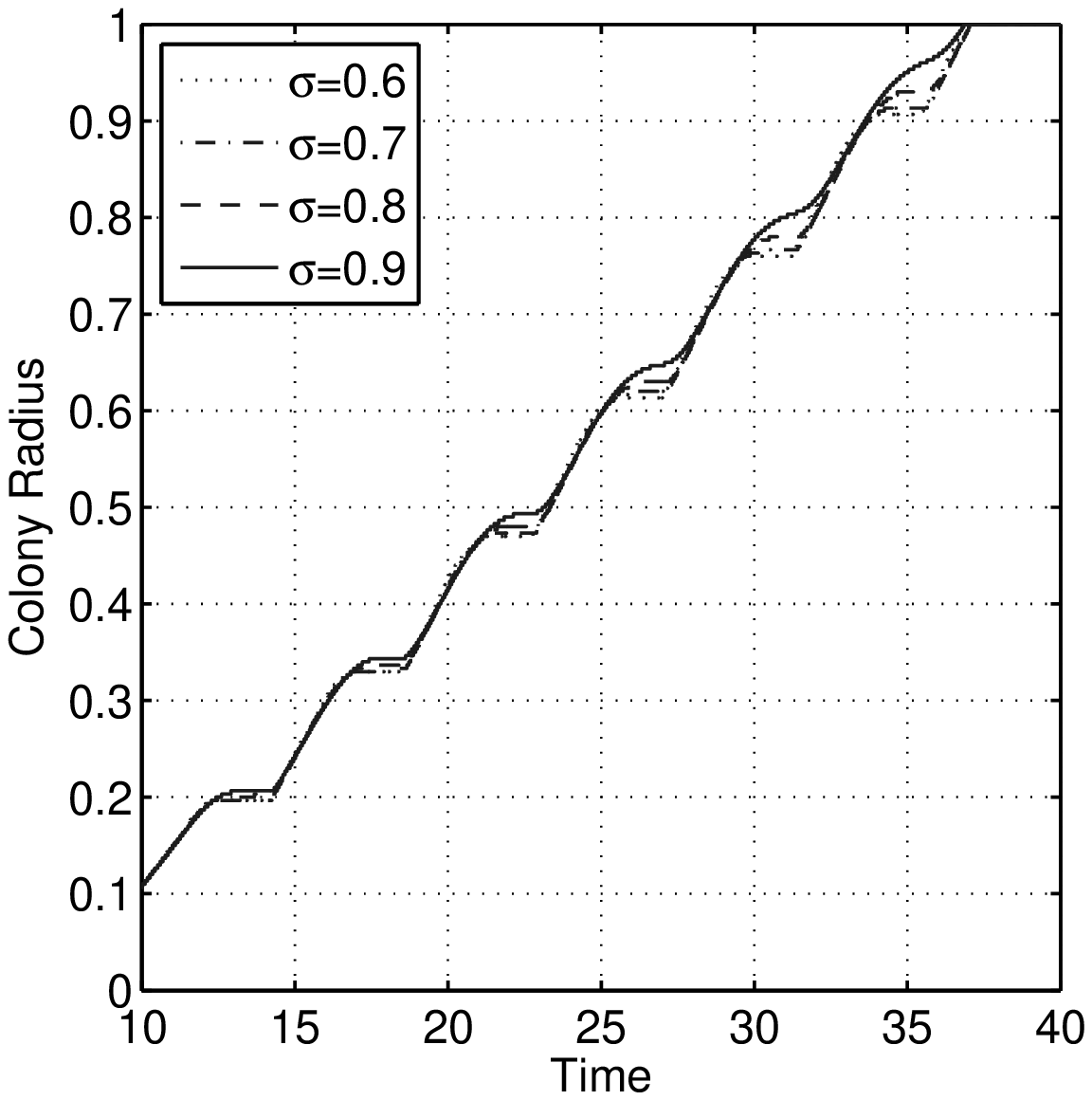,height=3in}
}
\caption{Swarm-colony front dynamics showing the breakdown of spatial and temporal regularity as the
  dedifferentiation age becomes less sharp.  Parameters common to all three swarm colonies are
  $\mu=1.0$, $\Amax = 2.67$, $v_c = 8.0$, $D_0 = 2\times 10^{-3}$,
  $\Pmin = 0.5$, $\xi_0 = 0.5$, and $\Amin=0$.}
\label{sigma}
\end{figure}

We next examine the response to changes in $\sigma$.  This parameter
study corresponds to the change in the probability function $\Pi(a)$
that a swarmer cell has not dedifferentiated by age $a$ shown in
figure \ref{PiHeavi}.  As illustrated in figure \ref{sigma}, colony
regularity breaks down as the age of dedifferentiation becomes less
sharp.  It is important to note that the colony remains
regular within an interval around $\sigma=0$.  As mentioned above, the
swarming cycle becoming irregular for $\sigma$ near zero, but not
equal to zero, would invalidate the model.  It is interesting to note that the breakdown of
colony regularity is nearly hysteretic as a function of $\sigma$.  In
the results shown in figure \ref{sigma}, we set $\mu_0=1$.  Results do
not change qualitatively for $\mu_0$ near this choice, however, as
shown in figure \ref{colonyPlots}, $\mu_0$ is an important parameter
in determining colony regularity.  Fortunately, we have the shape of
$\Pi(a)$ as a means to determine which parameter choices represent a sharp
versus a mollified age of dedifferentiation.

\begin{figure}[t]
\centering
\epsfig{file=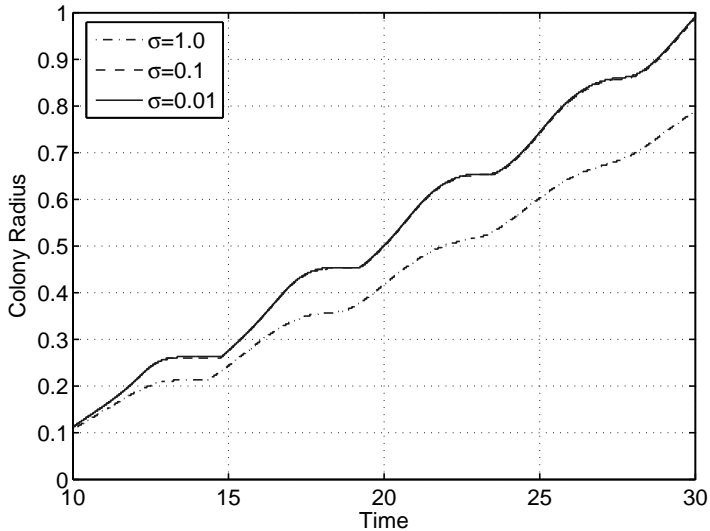,height=3in}
\caption{Swarm-colony front dynamics showing irregularity of all
  colonies when $\mu_0/\sigma$ is fixed.  For the colonies shown in
  this figure, $\mu_0 = \sigma$.  Parameters common to all three swarm colonies are
  $\Amax = 2.67$, $v_c = 8.0$, $D_0 = 2\times 10^{-3}$,
  $\Pmin = 0.5$, $\xi_0 = 0.5$, and $\Amin=0$.}
\label{cFixed}
\end{figure}

The last parameter exploration we present is that of fixing the ratio
$\mu_0/\sigma$ and varying $\sigma$.  This corresponds to the change in the probability function $\Pi(a)$
shown in figure \ref{PiDecay}.  As $\sigma \rightarrow 0$, the colony formation
never becomes truly regular.  This irregularity corresponds to a lack
of sharpness in $\Pi(a)$ as $\sigma \rightarrow 0$.  Figure
\ref{cFixed} illustrates our results for $\mu_0=\sigma$.  Setting
$\mu_0 = 3\sigma$ results in a loss of a swarming phase as $\sigma$
increases rather than the loss of a consolidation phase.

\section{Conclusions}
\label{conclusions}

This paper constitutes an argument that the regularity in the {\em Proteus} cell
cycle of differentiation and dedifferentiation between immotile
dividing cells and motile filament swarmer cells, along with
mechanisms that depend on cell density, underlies the
temporal and spatial regularity in swarm-colony development.  This
is at odds with the implications of reaction-diffusion models which attribute {\em Proteus} behavior
to density-dependent mechanisms alone.  The first conclusion to make
is that the question of the correct modeling approach might be
answered empirically by studying the colony regularity of strains of {\em Proteus} that do not
have a sharp age of dedifferentiation.   The probability that a
swarmer cell has not dedifferentiated by age $a$, denoted by $\Pi(a)$, can be used as
a guide to determine experimentally the forms of the dedifferentiation
modulus $\mu$ in equation (\ref{swarm}).

Density-dependent effects do matter in understanding {\em Proteus} swarm-colony
  development.  Experiments to determine how differentiation from dividing cells to
  swarmer cells depends on local population density would
  be particularly valuable in finding the forms of the
  differentiation function, $\xi$.

We, along with the other scientists modeling {\em Proteus} swarm-colony
development of whom we are aware, used diffusion to model movement.
Diffusion may not be the best representation of the spatial dynamics.  Areas of future research are the specifics of how the much
  larger and broader issue of biological motion relates to {\em
    Proteus} motion on an agar surface, and extending the models to
  explicit two-dimensional space to understand the interaction of
  multiple colonies.

We close by noting that the likely importance of the cell cycle in determining the macroscopic
behavior makes {\em Proteus} a model system for a multiscale
understanding of other biological systems where the spatial or temporal behavior is a manifestation
of local kinetics.  

\bibliographystyle{siam}
\bibliography{age,na,math,bio,forest,spatialecology}

\end{document}

%% file: trcover.tex
%
%
%
%

\voffset -.25in

\thispagestyle{empty}
\fontsize{12}{12}
\setcounter{page}{0}

\begin{center}
  \centerline{
  \psfig{figure=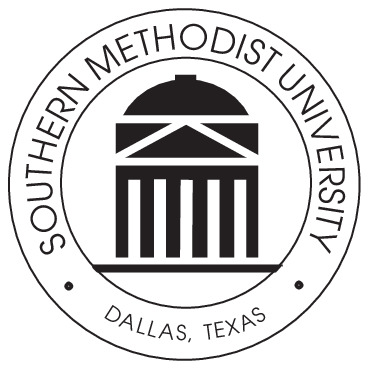,height=1.5in,width=1.5in}
  }

  \vspace{1.25in}

  \begin{minipage}[t]{4in}
    \begin{center}
  
      {\bf \trtitle

      \vspace{24pt}

      \trauths

      \vspace{12pt}

      SMU Math Report \trnum 
      }
    \end{center}
  \end{minipage}

  \vspace{3in}

  {\Large\bf D}{\large\bf EPARTMENT OF} 
  {\Large\bf M}{\large\bf ATHEMATICS} 

  \vspace{3pt}

  {\Large\bf S}{\large\bf OUTHERN} 
  {\Large\bf M}{\large\bf ETHODIST} 
  {\Large\bf U}{\large\bf NIVERSITY} 
\end{center}

\voffset 0in

\normalsize
\newpage